 \documentclass[sigconf]{acmart}
\pdfoutput=1

\AtBeginDocument{%
  \providecommand\BibTeX{{%
    \normalfont B\kern-0.5em{\scshape i\kern-0.25em b}\kern-0.8em\TeX}}}

\copyrightyear{2021}
\acmYear{2021}
\setcopyright{acmcopyright}\acmConference[ICMI '21]{Proceedings of the 2021 International Conference on Multimodal Interaction}{October 18--22, 2021}{Montréal, QC, Canada}
\acmBooktitle{Proceedings of the 2021 International Conference on Multimodal Interaction (ICMI '21), October 18--22, 2021, Montréal, QC, Canada}
\acmPrice{15.00}
\acmDOI{10.1145/3462244.3479889}
\acmISBN{978-1-4503-8481-0/21/10}

\acmSubmissionID{1234}

\begin{document}
\fancyhead{}
\title{To Rate or Not To Rate: Investigating Evaluation Methods for Generated Co-Speech Gestures}

\author{Pieter Wolfert}
\authornote{Corresponding Author}
\orcid{/0000-0002-7420-7181}
\affiliation{%
\institution{IDLab, Ghent University - imec}
\city{Ghent}
\country{Belgium}
}
\email{pieter.wolfert@ugent.be}

\author{Jeffrey M. Girard}
\orcid{0000-0002-7359-3746}
\affiliation{%
  \institution{Department of Psychology, University of Kansas}
  \city{Kansas}
  \country{USA}
}
\email{jmgirard@ku.edu}

\author{Taras Kucherenko}
\orcid{0000-0001-9838-8848}
\affiliation{%
  \institution{KTH Royal Institute of Technology}
  \city{Stockholm}
  \country{Sweden}
}
\email{tarask@kth.se}

\author{Tony Belpaeme}
\orcid{0000-0001-5207-7745}
\affiliation{%
\institution{IDLab, Ghent University - imec}
\city{Ghent}
\country{Belgium}
}
\email{tony.belpaeme@ugent.be}

\begin{abstract}
While automatic performance metrics are crucial for machine learning of artificial human-like behaviour, the gold  standard for evaluation remains human judgement. The subjective evaluation of artificial human-like behaviour in embodied conversational agents is however expensive and little is known about the quality of the data it returns. 
Two approaches to subjective evaluation can be largely distinguished, one relying on ratings, the other on pairwise comparisons. 
In this study we use co-speech gestures to compare the two against each other and answer questions about their appropriateness for evaluation of artificial behaviour. We consider their ability to rate quality, but also aspects pertaining to the effort of use and the time required to collect subjective data. 
We use crowd sourcing to rate the quality of co-speech gestures in avatars, assessing which method picks up more detail in subjective assessments. We compared gestures generated by three different machine learning models with various level of behavioural quality. 
We found that both approaches were able to rank the videos according to quality and that the ranking significantly correlated, showing that in terms of quality there is no preference of one method over the other. We also found that pairwise comparisons were slightly faster and came with improved inter-rater reliability, suggesting that for small-scale studies pairwise comparisons are to be favoured over ratings.
\end{abstract}



\maketitle

\section{Introduction}
When we interact with embodied conversational agents, we expect a similar manner of nonverbal communication as when interacting with humans. 
One way to achieve more human-like nonverbal behaviour in conversational agents is through the use of data-driven methods, which learn model parameters from data and gained in popularity over the past few years \cite{kucherenko2020gesticulator, ahuja-etal-2020-gestures, yoon2020speech, kucherenko2021large}. Data-driven methods have been used to generate lips synchronisation, eye gaze or facial expressions, however in this work we take co-speech gestures as a test bed for comparing evaluation methods. 
Data-driven methods are able to generate a wider range of gestures and behaviours, as behaviour is no longer restricted to pre-coded animation or procedurally generated behaviour, but instead are generated from models trained on large amounts of data of human movement. 
These behaviours are often used to drive conversational agents in both virtual and physical agents, as these improve interaction \cite{salem2013err, ham2015combining, chidambaram2012designing,lucca2018communicating,prieto2017beat}. 
Co-speech gestures are traditionally divided into four dimensions: iconic gestures, beat gestures, deictic gestures, and metaphoric gestures \cite{mcneill1992hand}. 
The approach to produce each of these categories often differs but, using data-driven methods, it becomes possible to generate multiple categories of gestures with a single model. 

The quality of generated human-like behaviour can be assessed using objective or subjective measures. 
\emph{Objective measures} rely on an algorithmic approach to return a quantitative measure of the quality of the behaviour and are entirely automated, while \emph{subjective measures} instead rely on ratings by human observers. 
Most recent papers on co-speech gesture generation report objective measures to assess the quality of the generated behaviour, with measures such as velocity diagrams or average jerk being popular \cite{kucherenko2021moving, ahuja-etal-2020-gestures, yoon2020speech}. 
These measures not only are easy to automate, but also allow comparisons across models. 
For example, Yoon et al. \cite{yoon2020speech} trained models from other authors on the same dataset to compare objective metric results. 
For this reason, objective measures are often preferred over subjective evaluations, as the latter are harder to compare due to their potentially high variability. 
Yet, subjective evaluations are crucial when evaluating the behaviour of agents interacting with humans. 
This is because social communication is much more complex than current objective measures are capable of capturing,  and subjective evaluations are still considered to be the gold standard. 
There may also be a large subjective component to how observers interpret generated behaviour, which we would like to capture. 
Thus, the ``final stretch'' in quality evaluations still relies heavily on subjective evaluations \cite{wagner2019speech}. 

While the value of subjective evaluations is widely accepted, there is little  consensus on how to collect and analyse such evaluations in relation to the evaluation of data-driven generated stimuli. 
Recently, several authors working on data-driven methods for nonverbal behaviour generation moved from rating scales (e.g., having observers rate how human-like two generated stimuli are from 1-to-5) to the use of pairwise comparisons (e.g., having observers select which of two generated stimuli is more human-like) \cite{kucherenko2020gesticulator, perez2019part, yoon2020speech, wolfert2019should}.
For example, Yoon et al. \cite{yoon2020speech} argued that \textit{``co-speech gestures are so subtle, so participants would have struggled to rate them on a five- or seven-point scale''} and promoted the use of pairwise comparisons over rating scales. 
Relatively little empirical attention has been devoted to this methodological topic in regard to the evaluation of data-driven generated stimuli, however, and it is still unknown how much the methods actually differ in terms of usability and informativeness.

In the current study, we seek to explore the similarities and differences between the rating scale and pairwise comparison approaches. 
We take generated co-speech gestures as a test bed for our evaluations but note that these findings may also apply to stimulus evaluation in other areas. 
Our hypotheses, design, and methodology were pre-registered before the data was gathered\footnote{\url{https://osf.io/7d9fs}}. 
We present short video clips to human participants, with each video clip showing an avatar displaying combined verbal and nonverbal behaviour. The movements are generated using three data-driven methods of varying quality and we expect the subjective evaluations to clearly reflect this difference. 
In order to gain more insight into the effectiveness of the two subjective evaluation methods, we formulated the following five hypotheses. 
\\
\begin{enumerate}
    \item[H1.] The rank-order of stimuli implied by the pairwise comparisons and rating scales will be different.
    \item[H2.] Pairwise comparisons will have higher inter-rater agreement than rating scales.
    \item[H3.] Pairwise comparisons and rating scales will differ in terms of time-efficiency (e.g., the time it takes for a single participant to finish a single evaluation).
    \item[H4.] Pairwise comparisons and rating scales will differ in terms of participant usage preference and usability (both qualitative and quantitative).
    \item[H5.] Pairwise comparisons and rating scales will both find a difference between stimuli that have a pronounced quality difference, but will not have enough resolution to find a difference between stimuli that differ slightly in quality.
\end{enumerate}
\vspace{0.5cm}

Our aim is to quantify the pros and cons of these subjective evaluation methods and to provide empirical recommendations for the community working on gesture generation on when to use each method. 
Although the concept of comparing these evaluation strategies is not novel on its own \cite{fisher1968comparison}, it is novel in relation to the evaluation of gesture generation for ECAs. We hope that this work can both highlight similarities and differences between the evaluated methods, and function as a bridge between the different fields of psycho-metrics and gesture generation researchers. 

\section{Related Work}
In this section, we cover work that compared rating scale evaluations with pairwise comparisons, and look at their specific use in the field of gesture generation for embodied conversational agents. 
To our knowledge, there has not been a comparison between rating scales and pairwise comparisons for the evaluation of co-speech gestures in Embodied Conversational Agents (ECAs) in particular. 
However, subjective evaluation methods have been studied and compared in several other fields.
We want to highlight that we believe that there is a difference in the type of stimuli that are evaluated: we consider data-driven behaviour in virtual agents. We are aware of the overlap there is with psychology and psychometrics, but want to zoom in specifically on the use of both rating and pairwise assessments of generated gesticular behaviour for ECAs. 

There is a rich history of work in psychology on related topics. 
DeCoster et al. \cite{decoster2009} compared analysing continuous variables directly with analysing them after dichotomisation (e.g., re-coding them as two-class variables such as high-or-low). 
Although there were a few edge cases where dichotomisation was similar to direct analysis, they demonstrated that dichotomisation throws away important information and concluded that the use of the original continuous variables is to be preferred in most circumstances.
Simms et al. \cite{simms2019} randomly assigned participants to complete the same personality rating scales with different numbers of response options ranging from two to eleven. 
They found that including four or fewer response options often attenuates psychometric precision, and including more than six response options generally provides no improvements in precision.
Finally, Rhemtulla et al. \cite{rhemtulla2012} demonstrated that treating rating scale data as continuous can be problematic (i.e., can result in biased estimates) for scales with fewer than five response options, which tend to be quite non-normally distributed.
Such data thus requires specialised ordinal methods to analyse properly.
Overall, the psychological literature thus suggests that rating scales with between five and seven response options would be preferable to rating scales with fewer response options.
If we consider the pairwise comparison approach to be similar to a rating scale with two response options (e.g., better or worse), this would raise concerns about the approach's psychometric precision and normality.

Although not covered in this paper, another way of evaluating stimuli on a continuous scale is by using visually-aided rating (VAR) \cite{janhunen2012comparison}. 
Visually, categories are still used as anchors in VAR, but specific scores are not visualised in comparison to Likert scales. 
This enables participants to quantify an ordering, from which it is still possible to derive a quantifiable rating. 
VAS-RRP is congruent to VAR, except that in VAR the rating scale is placed vertically, and in VAS-RRP horizontally \cite{sung2018visual}. 

However, there have also been impassioned arguments in favour of ordinal and rank-based approaches (of which the pairwise comparison approach can be considered a simple variant) within the affective computing community in recent years \cite{martinez2014, yannakakis2015,yannakakis2021}. 
The argument is that many subjective evaluations are inherently ordinal and cannot be adequately treated as continuous numbers or nominal categories and should instead be handled using rankings.
If this argument is accurate, then the pairwise comparison approach would be preferable to the rating scale approach on theoretical grounds.
There is also evidence that rank-based approaches might have some practical benefits over rating scale approaches, such as being faster to administer and more reliable over time.
For example, Clark et al. \cite{clark2018rate} evaluated the perception of physical strength from images of male bodies using both pairwise comparisons and rating scales and found that the scores were closely correlated but that the pairwise comparisons were completed 67\% faster. 
Other examples, like Elliot et al. \cite{elliott1958reliability} and Mueser et al.\cite{mueser1984you} found high correlations between rankings resulting from the evaluation of physical features in humans. 
Liang et al. \cite{liang2020beyond} proposes a model to `calibrate' self-reported user ratings for dialogue systems due to issues with validity and bias. 
In relation to biomedical image assessments, where evaluation considers the visual quality of the stimuli, Phelps et al. \cite{phelps2015pairwise} found that pairwise comparisons and ranked Likert scores made for more accurate assessments in comparison to the use of non-ranked Likert scores. 
Burton et al. \cite{burton2019best} compared rating scales with best-worst scaling, another variant of the rank-based approach. 
In this study, participants were asked to select the most attractive and least attractive faces in a series of images. 
The best-worst scaling approach showed better test-retest reliability than the rating scale approach.



One of the reasons the community would benefit from greater standardisation of subjective evaluations methods can be found in the recently organised GENEA gesture-generation challenge \cite{kucherenko2021large}. 
Invited researchers were asked to submit models trained on the same dataset containing human speech and co-speech gestures. 
All submissions were then compared using crowd-sourced subject evaluation, in which online participants were asked to rate each clip with a score between 1 and 100.
The benefit of this method is that generation models from different authors can be tested at once, within the same framework and participant pool. 
Sticking to a single evaluation strategy makes it possible to compare work across models, such as in the GENEA challenge, and also across time. 

Finally, a recently published preprint reviewed the literature on evaluation of gesture-generation systems \cite{wolfert2021review}. The review found that stimuli were often evaluated using very different rating scales, such as likeability, naturalness, and gesture-timing. 
This variability makes comparisons difficult across papers and time.

\section{Methods}
\begin{figure*}[h!]
\includegraphics[width=\linewidth]{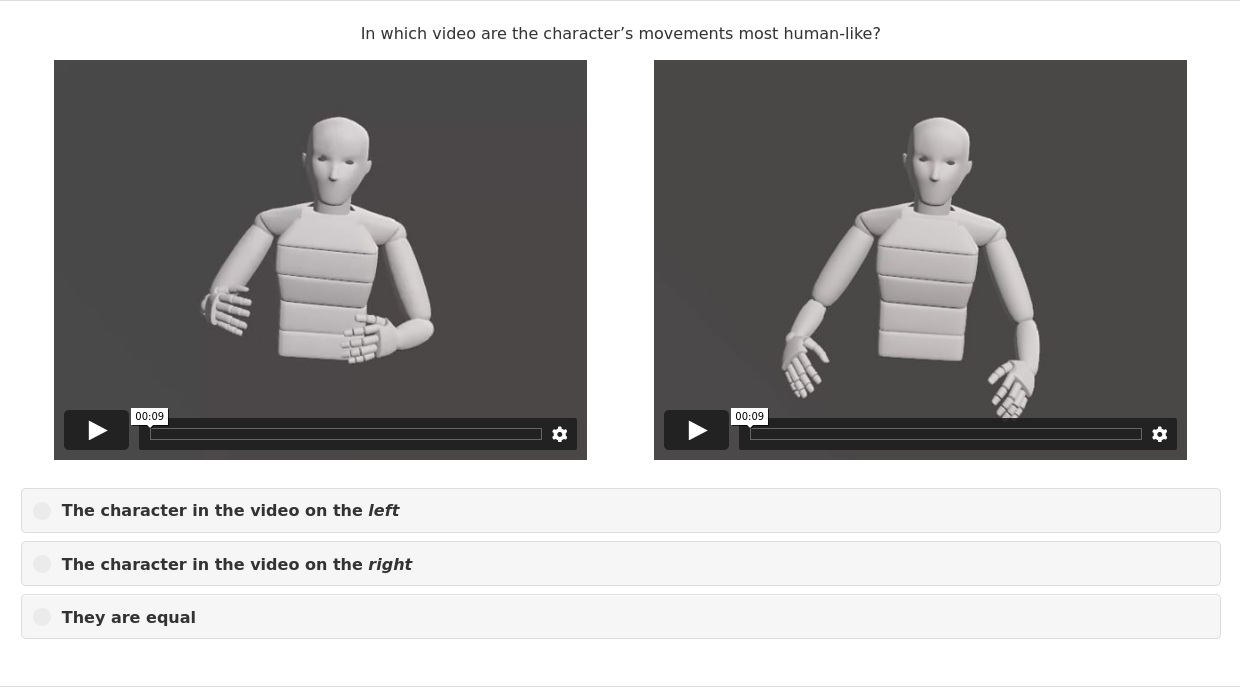}
\caption{Interface for pairwise comparison evaluation}
\label{fig:pwc}
\end{figure*}
\subsection{Experimental Design}
In this study, we used 30 video stimuli\footnote{\url{http://svito-zar.github.io/gesticulator/}} showing a gesticulating avatar provided by Kucherenko et al. \cite{kucherenko2020gesticulator}, the stimuli are already available and have been used by other researchers \cite{jonell2020can}. 
The videos had a resolution of $640\times480$ pixels and a frame rate of 30 frames per second. Three types of videos were used: Full, NoSpeech and NoText.
The \textit{Full} videos were generated by a model trained on motion of a human actor with the model having access to both the audio speech and transcribed text; the \textit{NoSpeech} videos were generated from a model only trained on motion and transcribed text; and the \textit{NoText} videos were generated by a model trained on motion and speech audio only. 
Thirty videos were created per type and, in each triplet of videos (across type), the avatar spoke the same sentence to facilitate comparison.
We have two study conditions: Full versus NoSpeech (which we denote Low Difference) and Full versus NoText (High Difference). We denote them this way because the former showed a small difference in the original study \cite{kucherenko2020gesticulator}, while the latter showed a large difference.
These conditions (Full. vs NoSpeech and Full. vs NoText) turned out to show significant differences in quality, and we assume that our subjective evaluations will reflect this.

Each participant in the current study was assigned to either the LowDiff or HighDiff condition. Following that, the participant was assigned to one of two ordering conditions:

\begin{enumerate}
\item [PR:] Pairwise Comparison approach for 10 videos drawn from a set of 30 videos, followed by the Rating Scale approach for the same 10 videos.
\item [RP:] Rating Scale approach for 10 videos drawn from a set of 30, and then Pairwise Comparison approach for the same 10 videos. 
\end{enumerate}

\subsection{Participants}
For this study, 130 participants were recruited on Prolific\footnote{https://www.prolific.co/}. 
To ensure data quality, participants have to be a native speaker of English, have at least a 90\% approval rating on the platform, and have participated in at least 100 other studies on the platform.
Participants were assigned to conditions using block randomisation in order to maintain balanced conditions. 

\subsection{Technical Setup}
\begin{figure}[h]
    \centering
    \includegraphics[width=\linewidth]{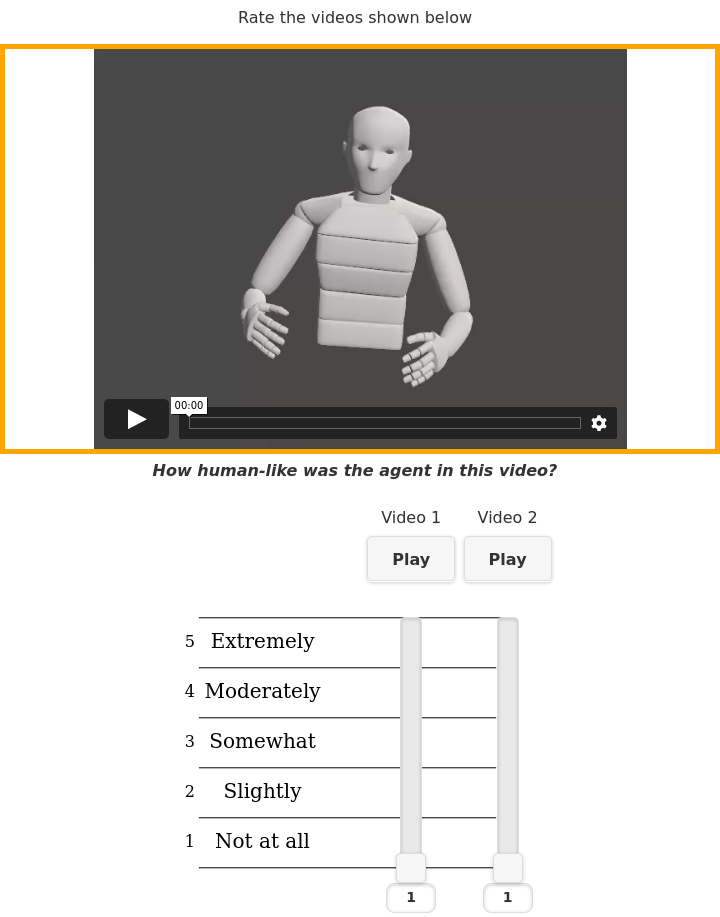}
    \caption{Interface for rating scale evaluation}
    \label{fig:rating}
\end{figure}
From Prolific, participants were forwarded to a web application to evaluate the stimuli. 
This application was based on HEMVIP \cite{jonell2021hemvip}, which in turn was based on WebMushra \cite{schoeffler2018webmushra} but adapted to work with video files.
Since two evaluation strategies were evaluated, there were two interface versions. 

The pairwise comparison interface (\autoref{fig:pwc}) displays two videos side by side, with three options for evaluation displayed below the videos. 
For all conditions, the question was: `In which video are the character's movements most human-like?'
The three response options were: \textit{left}, \textit{right}, and \textit{equal}. 
Participants were able to play both videos at the same time, but it is not explicitly mentioned in the instructions. 
After the participants watched both videos and selected a response option, they could continue to the next page.

The rating scale interface (\autoref{fig:rating}) displays a single video at a time, with a rating scale displayed below. 
For all conditions, the question was: `How human-like was the agent in this video?'
Response options ranged from 1 to 5 and were labelled \textit{not at all}, \textit{slightly}, \textit{somewhat}, \textit{moderately}, and \textit{extremely}.
Videos could only be watched one-at-a-time, and participants were only able to advance to the next page when both videos had been played and rated.

\subsection{Experimental Procedure}
After participants were assigned to the task on Prolific, they were forwarded to the online evaluation system. 
Here, they were assigned an internal participant ID that corresponds to a configuration file containing the stimuli and order of stimuli to show to the participant, and when to run attention checks.
Each participant evaluated a total of 22 video pairs. 
These 22 video pairs correspond to 10 videos evaluated in a pairwise comparison approach, and 10 in a rating style approach. Two of the 22 video pairs contained an attention check. 
The order of evaluation (pairwise comparison vs. rating approach) was based on the assigned ordering condition.
The position of the attention checks in the series of evaluation pairs was randomised, and there were two types of attention checks: one in which the response option to select was provided visually and one in which it was provided acoustically.

After evaluating the 22 video pairs, participants were presented with a questionnaire collecting their age, gender, nationality, level of education and experience with computers. 
This was followed with open questions related to the procedure they just completed, and whether they had a preference for pairwise comparison or rating scale evaluations. 
Once done with the study, successful participants were rewarded with 2.50 GBP (pay on average was 7.23 GBP per hour when taking into account the average duration of the task). The time each participant spent on each page of the experiment (and overall) was also recorded to allow us to evaluate efficiency.

\section{Analyses}
\subsection{Hypothesis 1}
To test the hypothesis that the two comparison methods would result in different rank-orderings of stimuli, we used a correlational approach. 
We first calculated each stimulus' average score across participants for each comparison method. 
Average scores using the rating scale method ranged from 1 to 5, and average scores for the pairwise approach ranged from --1 to 1 (on a scale where $1$ = the stimulus was preferred over the alternative, $0$ = the stimulus and alternative were equal, and $-1$ = the alternative was preferred over the stimulus). 
We then estimated the Kendall Rank-Order correlation \cite{kendall1938} between these two series.

\subsection{Hypothesis 2}
To test the hypothesis that the pairwise comparison method would have higher inter-rater agreement than the rating scale method, we used two statistical approaches. 
First, we estimated intraclass correlation coefficients (ICCs) using Model 2A \cite{mcgraw1996} and calculated the absolute agreement of the average of 12 participants (i.e., the minimum number of participants assigned to any comparison).
This approach estimates the reliability of the average of multiple participants' responses (which is what is used to compare video-generating methods), but assumes that the data approximates a continuous distribution (which is not the case for the pairwise method). 
As such, we also estimated chance-adjusted categorical agreement using quadratic-weighted kappa coefficients \cite{gwet2014}. 
This approach is overly pessimistic in this case because it estimates the reliability of a single randomly selected participant's response, but it has the benefit of not assuming continuous data.
In both cases, 2000 iterations of non-parametric bootstrapping \cite{Efron1993} (with percentile-based confidence intervals and $p$-values) were used to compare the two approaches' inter-rater reliability.

\subsection{Hypothesis 3}
To test the hypothesis that the two comparison methods would differ in terms of time-efficiency (i.e., the time it takes a participant to complete a single comparison/page), we used a linear mixed effects modelling approach \cite{galecki2013linear}. 
We estimated a model in which each page's completion time (in seconds) was regressed on a binary variable representing the comparison method. 
To control for practice and fatigue effects, we also regressed the completion time variable on a binary variable representing whether the comparison was during the first or second half of the experiment, and the method-by-half interaction effect to allow the difference between comparison methods to differ between the first and second half of the experiment. 
Finally, to account for the clustering/nesting of comparisons within participants and videos, we included random intercepts for these variables and used Satterthwaite's approximation \cite{kuznetsova2017} to correct model degrees of freedom for small clusters.

\subsection{Hypothesis 4}
To test the hypothesis that participants would be more likely to prefer the pairwise comparison approach than the rating approach, we estimated an intercept-only logistic regression model to predict a binary variable representing whether each participant preferred the pairwise comparison approach over the rating comparison approach. 
We then back-transformed the intercept to probability units and tested whether it was significantly different from an equal preference of 50\%.

\subsection{Hypothesis 5}
To test the hypothesis that the two comparison methods (i.e. rating scale and pairwise) would both find a difference in the case of a large difference in the quality of generated behaviour (i.e., Full vs. NoText stimuli) but not in the case of a small difference in the quality of generated behaviour (i.e. Full vs. NoSpeech stimuli), we used a linear mixed effects modelling approach \cite{galecki2013linear}. 
We estimated a model in which the choice for the Full stimuli was regressed on other (NoText or NoSpeech) and order. 

\section{Results}
130 participants were recruited, of which 100 participants passed the attention checks. Of these, the mean age was 35.01 (SD=12.64), 55 identified as female, 45 as male. 
68 of the participants were UK nationals, 22 were from the USA, 4 participants were Canadian, 2 Irish, 1 Australian, 1 Bulgarian, 1 Indian and 1 from New Zealand. 

\subsection{Hypotheses}
\subsubsection{Hypothesis 1}
\begin{figure}[h]
    \centering
    \includegraphics[width=\linewidth]{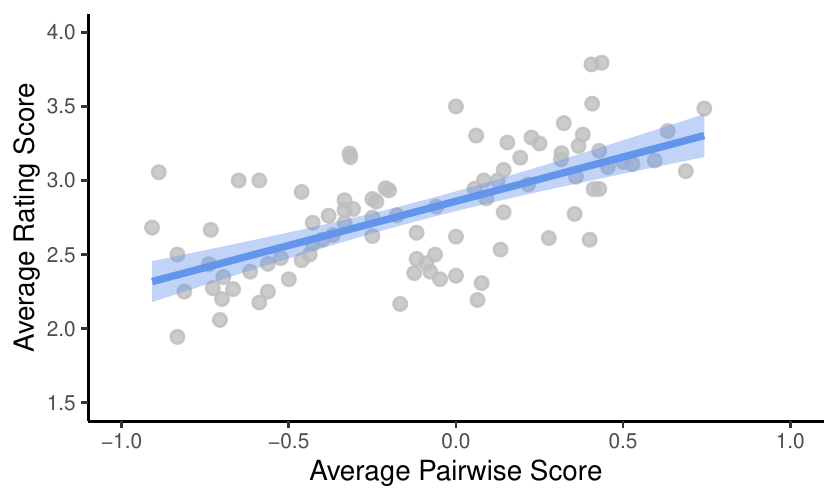}
    \caption{Relationship between average rating and pairwise scores. The two are positively correlated.}
    \label{fig:ratingvspairwise}
\end{figure}

In \autoref{fig:ratingvspairwise}, we can see the relationship between the average pairwise scores and the average rating scores. 
We quantified the magnitude of this relationship using Kendall's Rank-Order Correlation. 
When we excluded trials where the two stimuli being compared were rated as equally human-like, we found a rank correlation of 0.44, 95\% CI: [0.32, 0.55], $p<.001$. 
When we included trials where the two stimuli being compared were rated as equally human-like and assigned a pairwise score of 0, this correlation became 0.46, 95\% CI: [0.35, 0.57], $p<.001$. 
Thus, although the two methods did not have exactly the same rank-ordering of stimuli, their rank-orderings were positively correlated (i.e., similar) to a high degree.

\subsubsection{Hypothesis 2}
Using the intraclass correlation approach, the inter-rater reliability coefficient was 0.62, 95\% CI: $[0.50, 0.69]$ for the rating scale method and 0.77, 95\% CI: $[0.71, 0.82]$ for the pairwise method; this difference was statistically significant $(\Delta=0.15$, 95\% CI: $[0.06, 0.27]$, $p<.001)$.
Using the chance-adjusted categorical agreement approach, the quadratic-weighted kappa coefficient was 0.14, 95\% CI: $[0.09, 0.18]$ for the rating scale method and 0.23, 95\% CI: $[0.18, 0.28]$ for the pairwise method; this difference was statistically significant $(\Delta=0.09$, 95\% CI: $[0.03, 0.16]$, $p=.009)$.

\subsubsection{Hypothesis 3}
\begin{figure}[h]
    \centering
    \includegraphics[width=\linewidth]{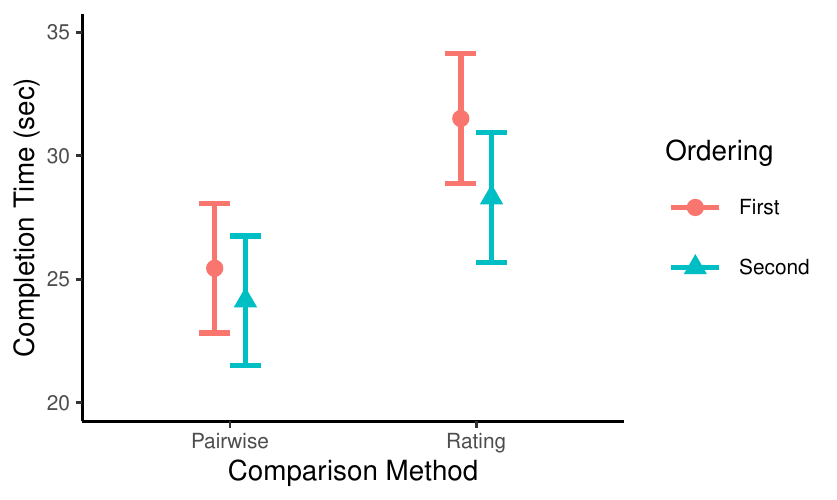}
    \caption{Completion time across conditions (error bars are 95\% CIs), showing that the pairwise method is approximately 6 seconds faster per page than the rating method.}
    \label{fig:timeorder}
\end{figure}

The main effect of comparison method was significantly greater than zero, $B=6.07$, 95\% CI: $[2.36, 9.77]$, $p=.002$ (see \autoref{fig:timeorder}). 
The unstandardised slope estimate of 6.07 means that pages were completed an average of around 6 seconds faster for the pairwise approach than for the rating approach. 
The main effect of ordering was not significantly different from zero $(p=.491)$ and the type-by-ordering interaction effect was also not significantly different from zero $(p=.600)$, which means that completion time did not significantly differ between the first and second half of the experiment and that the difference between comparison methods did not depend on which came first or second in the experiment.

If we want to know what the time difference would be for an entire experiment, we can multiply this page-level effect by the number of pages shown to participants. 
For 10 pages, as we did in this study, the experiment-level difference would be around 60 seconds.

\subsubsection{Hypothesis 4}
The intercept for preference for the pairwise method was estimated at 56.0\%, 95\% CI: $[46.2\%, 65.5\%]$ and was not significantly different from an equal preference of 50\% $(p=.231)$. Thus, we cannot conclude that participants reliably preferred one method over the other.


\subsubsection{Hypothesis 5}
\begin{figure*}[h!]
    \centering
    \includegraphics[width=\linewidth]{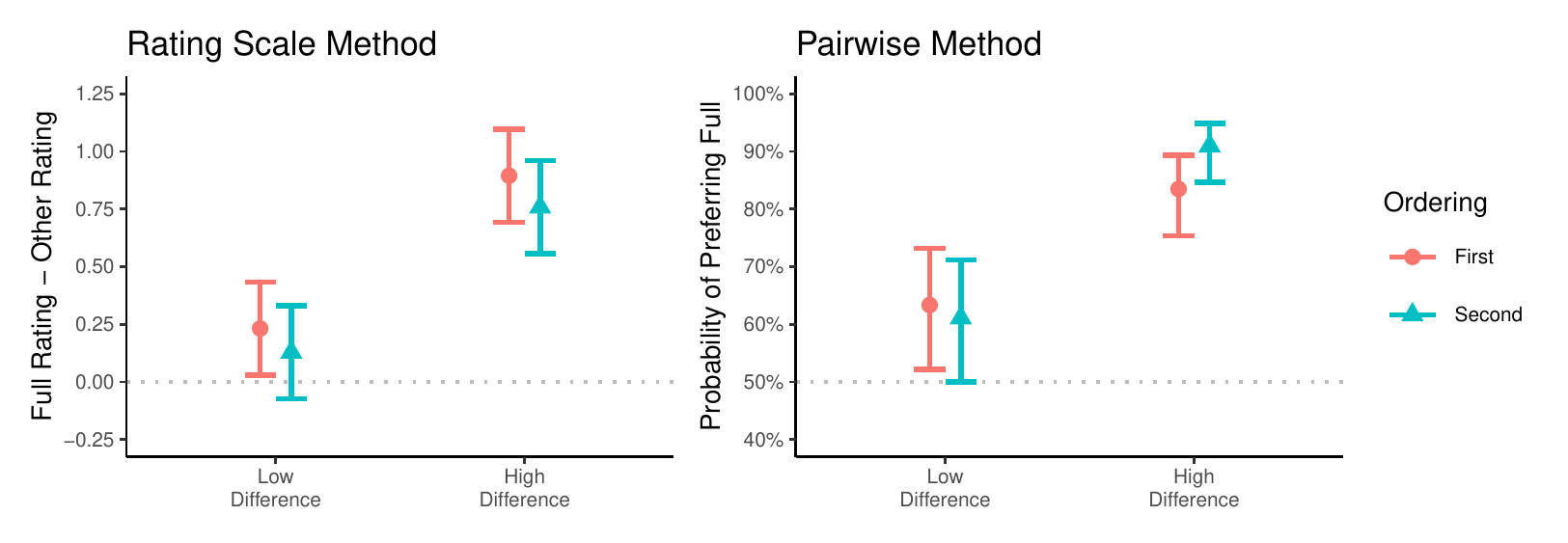}
    \caption{Comparison of generation methods by condition and evaluation method (error bars are 95\% CIs)}
    \label{fig:difference}
\end{figure*}
For the rating scale method, the main effect of other was significantly greater than zero, $B=0.66$, 95\% CI: $[0.40, 0.92]$, $p<.001$.
This means that the extent to which the Full stimuli were rated higher than the other stimuli was greater for the HighDiff stimuli than for the LowDiff stimuli. In this model, neither the main effect of ordering ($p=.439$) nor the other-by-ordering interaction effect ($p=.860$) were significant.  
For the pairwise method, the main effect of other was significantly greater than zero, $B=1.07$, 95\% CI: $[0.45, 1.69]$, $p<.001$. 
This means that the probability of preferring the full stimulus over the other stimulus was greater for the HighDiff stimuli than for the LowDiff stimuli. In this model, neither the main effect of ordering ($p=.750$) nor the other-by-ordering interaction effect ($p=.094$) were significant. Despite different scaling, the two methods had very similar results that matched our hypotheses and also matched the results from the original study we were reproducing \cite{kucherenko2020gesticulator} (see \autoref{fig:difference}).

\section{Discussion}
In this study, we explored the differences in evaluating gesture motion stimuli with both pairwise comparisons and rating scales. 
Our aim was to gain a deeper understanding of when to use each approach. 
For this, we looked at the stimulus rankings both methods provided, their inter-rater reliability, the time it took participants to complete evaluations, participant preferences, and the conclusions both methods would yield regarding the comparison of gesture generation methods with high and low differences in quality. 

The rank-ordering of stimuli between the pairwise comparisons and rating scales had a moderate positive correlation. We can conclude that in order to rank stimuli, in this instance co-speech gestures, there is not one approach that is preferred over the other; both are able to subjectively distinguish bad from good stimuli and this can be used to establish an order of quality.

When we take a look at the inter-rater reliability, we see a higher reliability for the pairwise method. 
This suggests that the pairwise method might be preferred over the rating scale method in terms of reliability. 

When we look at which approach is faster, we can conclude that each comparison using the pairwise method was, on average, 6 seconds faster (25s instead of 31s) than each comparison using the rating scale method, which aligns with the findings of previous studies \cite{clark2018rate}.
Although this difference was statistically significant (i.e., reliable), a difference of 6 seconds per comparison is likely too small to make much of a practical difference unless the number of comparisons being made by each participant was large (e.g., 100 or more).

Whether participants reliably preferred one comparison method over the other depended on which method they were assigned to use first. 
Those participants who used the rating scale method and then the pairwise method significantly preferred the pairwise method. 
However, those who used the pairwise method and then the rating scale method did not show a reliable preference for either method. 
This provides tentative evidence that the pairwise method may be more user-friendly.

In line with a previous study \cite{kucherenko2020gesticulator}, we found that a high qualitative difference is indeed picked up by subjective evaluations. 
Not only does this hold for pairwise comparisons, but also for the rating scale approach. 
Both methods can provide similar results and are equivalent when comparing two or more conditions, for example two different models used to generate behaviour. 

\subsection{Limitations}
For this study we gathered 2200 evaluations submitted by 100 participants. Due to the random drawing of stimuli, stimuli did not have the same number of responses. A pseudo-random spreading of stimuli over participants could have avoided this. The fact that participants could watch the videos simultaneously in the pairwise comparison interface but not in the rating scale interface may have contributed to the difference in average completion time between the two methods.
We only considered `human-likeness' in terms of assessing the quality of the generated gestures assessed by the participants, and are aware how limiting this question is in relation to the full spectrum of possible questions in relation to the evaluation of these stimuli. 
We opted for this strategy as the aim of this work was not to demonstrate which questions are most appropriate for the evaluation, but to compare the outcomes of two different evaluation strategies. 

\subsection{Recommendations}

Based on our results, we have found no strong evidence to prefer one evaluation method over the other. The study does however allow us to make a number of recommendations for each method in relation to the domain of gesture generation, taking into account previous studies in other domains. 

\emph{Pairwise comparisons} may be better suited when a large number of stimuli are to be evaluated, as this not only results in a shorter study but is likely to avoid fatigue in participants. If only a small number of conditions are under consideration, then pairwise comparisons of conditions is practical, but as the number of combinations grows with the faculty of the number of conditions $(\frac{n!}{2(n-2)!}$, with $n$ the number of conditions) pairwise comparisons tend to become unwieldy for 4 or more conditions if we want to compare all versus all.

\emph{Rating scales} may be more appropriate when fine-grained evaluations are needed, as ratings can not only be used to rate stimuli between conditions, but can also be used to rank stimuli within conditions. Ratings are also recommended when more than 3 conditions are under considerations, as the number of required ratings grows linearly with the number of conditions and stimuli. We would however like to emphasise the importance of providing anchors/labels for each response option in the rating scales \cite{weijters2010effect}. When using rating scales, it is also recommended to calibrate participants' judgements by showing the participants poor and excellent stimuli during a brief training session. While the lack of calibration can somewhat be addressed by normalising participants' ratings, resolution and reliability are lost when participants are not properly trained before starting their rating task.

Finally, it is important to consider the type of information provided by each evaluation method. 
Rating scales provide information about the quality of each stimulus on an \textit{absolute} scale, whereas pairwise comparisons provide information on a \textit{relative} scale. 
Thus, you could use the pairwise comparison method to establish whether one method of generating human-like behaviour was reliably preferred over another. 
However, being `better' is not always the same as being `good'. 
For instance, one method could be considered `poor' and the other `very poor'; this would likely result in a big difference in pairwise comparisons, but it would be a mistake to conclude that the former was therefore high quality in absolute terms. 
This is where carefully crafted rating scales (and qualitative methods, such as interviews and free response boxes) can provide additional information about quality in general.

\section{Conclusion}
Objective evaluation measures of generated human-like behaviour often provide insufficient information to fully assess the quality of the behaviour. As such, these measures are often supplemented with subjective evaluations. However, the field of gesture generation currently overlooks the amount of work that has been done in other fields that deal with subjective evaluations. This paper compared two popular methods, pairwise comparisons and rating scales, and found that both were equally effective to assess the quality of generated behaviour and provided surprisingly similar results in terms of rank-ordering of stimuli, inter-rater reliability, participant usability preferences, and the conclusions they yielded regarding the comparison of different stimuli-generation methods.  We found that pairwise comparisons were slightly faster and showed somewhat higher inter-rater reliability, whereas the rating scale approach provided information on both absolute and relative quality and is capable of scales better when comparing more than two stimuli at a time. These insights are increasingly relevant in a time where quantitative quality measures are used to drive research and development, especially when the use of data-driven methods tends to put draw attention away from subjective measures in favour of objective loss functions. Subjective measures are likely to remain the gold standard in evaluation studies 
and a better understanding of their capabilities benefits the study of multimodal behaviour generation.

\section{Acknowledgment}
This research received funding from the Flemish Government (AI Research Program), was supported by the Flemish Research Foundation grant no. 1S95020N and the Swedish Foundation for Strategic Research contract no. RIT15-0107 (EACare). 

\balance

\bibliographystyle{ACM-Reference-Format}
\bibliography{ref}


\begin{thebibliography}{40}


\ifx \showCODEN    \undefined \def \showCODEN     #1{\unskip}     \fi
\ifx \showDOI      \undefined \def \showDOI       #1{#1}\fi
\ifx \showISBNx    \undefined \def \showISBNx     #1{\unskip}     \fi
\ifx \showISBNxiii \undefined \def \showISBNxiii  #1{\unskip}     \fi
\ifx \showISSN     \undefined \def \showISSN      #1{\unskip}     \fi
\ifx \showLCCN     \undefined \def \showLCCN      #1{\unskip}     \fi
\ifx \shownote     \undefined \def \shownote      #1{#1}          \fi
\ifx \showarticletitle \undefined \def \showarticletitle #1{#1}   \fi
\ifx \showURL      \undefined \def \showURL       {\relax}        \fi
\providecommand\bibfield[2]{#2}
\providecommand\bibinfo[2]{#2}
\providecommand\natexlab[1]{#1}
\providecommand\showeprint[2][]{arXiv:#2}

\bibitem[\protect\citeauthoryear{Ahuja, Lee, Ishii, and Morency}{Ahuja
  et~al\mbox{.}}{2020}]%
        {ahuja-etal-2020-gestures}
\bibfield{author}{\bibinfo{person}{Chaitanya Ahuja}, \bibinfo{person}{Dong~Won
  Lee}, \bibinfo{person}{Ryo Ishii}, {and} \bibinfo{person}{Louis-Philippe
  Morency}.} \bibinfo{year}{2020}\natexlab{}.
\newblock \showarticletitle{No Gestures Left Behind: Learning Relationships
  between Spoken Language and Freeform Gestures}. In
  \bibinfo{booktitle}{\emph{Findings of the Association for Computational
  Linguistics: EMNLP 2020}}. \bibinfo{publisher}{Association for Computational
  Linguistics}.
\newblock
\urldef\tempurl%
\url{https://doi.org/10.18653/v1/2020.findings-emnlp.170}
\showDOI{\tempurl}


\bibitem[\protect\citeauthoryear{Burton, Burton, Rigby, Sutherland, and
  Rhodes}{Burton et~al\mbox{.}}{2019}]%
        {burton2019best}
\bibfield{author}{\bibinfo{person}{Nichola Burton}, \bibinfo{person}{Michael
  Burton}, \bibinfo{person}{Dan Rigby}, \bibinfo{person}{Clare~AM Sutherland},
  {and} \bibinfo{person}{Gillian Rhodes}.} \bibinfo{year}{2019}\natexlab{}.
\newblock \showarticletitle{Best-worst scaling improves measurement of first
  impressions}.
\newblock \bibinfo{journal}{\emph{Cognitive research: principles and
  implications}} \bibinfo{volume}{4}, \bibinfo{number}{1}
  (\bibinfo{year}{2019}), \bibinfo{pages}{1--10}.
\newblock


\bibitem[\protect\citeauthoryear{Chidambaram, Chiang, and Mutlu}{Chidambaram
  et~al\mbox{.}}{2012}]%
        {chidambaram2012designing}
\bibfield{author}{\bibinfo{person}{Vijay Chidambaram},
  \bibinfo{person}{Yueh-Hsuan Chiang}, {and} \bibinfo{person}{Bilge Mutlu}.}
  \bibinfo{year}{2012}\natexlab{}.
\newblock \showarticletitle{Designing persuasive robots: how robots might
  persuade people using vocal and nonverbal cues}. In
  \bibinfo{booktitle}{\emph{Proceedings of the seventh annual ACM/IEEE
  international conference on Human-Robot Interaction}}.
  \bibinfo{pages}{293--300}.
\newblock


\bibitem[\protect\citeauthoryear{Clark, Howard, Woods, Penton-Voak, and
  Neumann}{Clark et~al\mbox{.}}{2018}]%
        {clark2018rate}
\bibfield{author}{\bibinfo{person}{Andrew~P Clark}, \bibinfo{person}{Kate~L
  Howard}, \bibinfo{person}{Andy~T Woods}, \bibinfo{person}{Ian~S Penton-Voak},
  {and} \bibinfo{person}{Christof Neumann}.} \bibinfo{year}{2018}\natexlab{}.
\newblock \showarticletitle{Why rate when you could compare? Using the
  “EloChoice” package to assess pairwise comparisons of perceived physical
  strength}.
\newblock \bibinfo{journal}{\emph{PloS one}} \bibinfo{volume}{13},
  \bibinfo{number}{1} (\bibinfo{year}{2018}), \bibinfo{pages}{e0190393}.
\newblock


\bibitem[\protect\citeauthoryear{DeCoster, Iselin, and Gallucci}{DeCoster
  et~al\mbox{.}}{2009}]%
        {decoster2009}
\bibfield{author}{\bibinfo{person}{Jamie DeCoster},
  \bibinfo{person}{Anne-Marie~R. Iselin}, {and} \bibinfo{person}{Marcello
  Gallucci}.} \bibinfo{year}{2009}\natexlab{}.
\newblock \showarticletitle{A Conceptual and Empirical Examination of
  Justifications for Dichotomization}.
\newblock \bibinfo{journal}{\emph{Psychological Methods}} \bibinfo{volume}{14},
  \bibinfo{number}{4} (\bibinfo{year}{2009}), \bibinfo{pages}{349--366}.
\newblock
\urldef\tempurl%
\url{https://doi.org/10/bh86w7}
\showDOI{\tempurl}


\bibitem[\protect\citeauthoryear{Efron and Tibshirani}{Efron and
  Tibshirani}{1993}]%
        {Efron1993}
\bibfield{author}{\bibinfo{person}{Bradley Efron} {and}
  \bibinfo{person}{Robert~J. Tibshirani}.} \bibinfo{year}{1993}\natexlab{}.
\newblock \bibinfo{booktitle}{\emph{An Introduction to the Bootstrap}}.
\newblock \bibinfo{publisher}{{Chapman and Hall}}, \bibinfo{address}{{New York,
  NY}}.
\newblock


\bibitem[\protect\citeauthoryear{Elliott}{Elliott}{1958}]%
        {elliott1958reliability}
\bibfield{author}{\bibinfo{person}{Lois~Lawrence Elliott}.}
  \bibinfo{year}{1958}\natexlab{}.
\newblock \showarticletitle{Reliability of judgments of figural complexity.}
\newblock \bibinfo{journal}{\emph{Journal of experimental psychology}}
  \bibinfo{volume}{56}, \bibinfo{number}{4} (\bibinfo{year}{1958}),
  \bibinfo{pages}{335}.
\newblock


\bibitem[\protect\citeauthoryear{Fisher, Weiss, and Dawis}{Fisher
  et~al\mbox{.}}{1968}]%
        {fisher1968comparison}
\bibfield{author}{\bibinfo{person}{Stephen~T Fisher}, \bibinfo{person}{David~J
  Weiss}, {and} \bibinfo{person}{Ren{\'e}~V Dawis}.}
  \bibinfo{year}{1968}\natexlab{}.
\newblock \showarticletitle{A comparison of likert and pair comparisons
  techniques in multivariate attitude scaling}.
\newblock \bibinfo{journal}{\emph{Educational and Psychological Measurement}}
  \bibinfo{volume}{28}, \bibinfo{number}{1} (\bibinfo{year}{1968}),
  \bibinfo{pages}{81--94}.
\newblock


\bibitem[\protect\citeauthoryear{Ga{\l}ecki and Burzykowski}{Ga{\l}ecki and
  Burzykowski}{2013}]%
        {galecki2013linear}
\bibfield{author}{\bibinfo{person}{Andrzej Ga{\l}ecki} {and}
  \bibinfo{person}{Tomasz Burzykowski}.} \bibinfo{year}{2013}\natexlab{}.
\newblock \showarticletitle{Linear mixed-effects model}.
\newblock In \bibinfo{booktitle}{\emph{Linear Mixed-Effects Models Using R}}.
  \bibinfo{publisher}{Springer}, \bibinfo{pages}{245--273}.
\newblock


\bibitem[\protect\citeauthoryear{Gwet}{Gwet}{2014}]%
        {gwet2014}
\bibfield{author}{\bibinfo{person}{Kilem~Li Gwet}.}
  \bibinfo{year}{2014}\natexlab{}.
\newblock \bibinfo{booktitle}{\emph{Handbook of Inter-Rater Reliability:
  {{The}} Definitive Guide to Measuring the Extent of Agreement among Raters}
  (\bibinfo{edition}{fourth} ed.)}.
\newblock \bibinfo{publisher}{{Advanced Analytics}},
  \bibinfo{address}{{Gaithersburg, MD}}.
\newblock


\bibitem[\protect\citeauthoryear{Ham, Cuijpers, and Cabibihan}{Ham
  et~al\mbox{.}}{2015}]%
        {ham2015combining}
\bibfield{author}{\bibinfo{person}{Jaap Ham}, \bibinfo{person}{Raymond~H
  Cuijpers}, {and} \bibinfo{person}{John-John Cabibihan}.}
  \bibinfo{year}{2015}\natexlab{}.
\newblock \showarticletitle{Combining robotic persuasive strategies: the
  persuasive power of a storytelling robot that uses gazing and gestures}.
\newblock \bibinfo{journal}{\emph{International Journal of Social Robotics}}
  \bibinfo{volume}{7}, \bibinfo{number}{4} (\bibinfo{year}{2015}),
  \bibinfo{pages}{479--487}.
\newblock


\bibitem[\protect\citeauthoryear{Janhunen}{Janhunen}{2012}]%
        {janhunen2012comparison}
\bibfield{author}{\bibinfo{person}{Kristiina Janhunen}.}
  \bibinfo{year}{2012}\natexlab{}.
\newblock \showarticletitle{A comparison of Likert-type rating and
  visually-aided rating in a simple moral judgment experiment}.
\newblock \bibinfo{journal}{\emph{Quality \& Quantity}} \bibinfo{volume}{46},
  \bibinfo{number}{5} (\bibinfo{year}{2012}), \bibinfo{pages}{1471--1477}.
\newblock


\bibitem[\protect\citeauthoryear{Jonell, Kucherenko, Torre, and Beskow}{Jonell
  et~al\mbox{.}}{2020}]%
        {jonell2020can}
\bibfield{author}{\bibinfo{person}{Patrik Jonell}, \bibinfo{person}{Taras
  Kucherenko}, \bibinfo{person}{Ilaria Torre}, {and} \bibinfo{person}{Jonas
  Beskow}.} \bibinfo{year}{2020}\natexlab{}.
\newblock \showarticletitle{Can we trust online crowdworkers? Comparing online
  and offline participants in a preference test of virtual agents}. In
  \bibinfo{booktitle}{\emph{Proceedings of the 20th ACM International
  Conference on Intelligent Virtual Agents}}. \bibinfo{pages}{1--8}.
\newblock


\bibitem[\protect\citeauthoryear{Jonell, Yoon, Wolfert, Kucherenko, and
  Henter}{Jonell et~al\mbox{.}}{2021}]%
        {jonell2021hemvip}
\bibfield{author}{\bibinfo{person}{Patrik Jonell}, \bibinfo{person}{Youngwoo
  Yoon}, \bibinfo{person}{Pieter Wolfert}, \bibinfo{person}{Taras Kucherenko},
  {and} \bibinfo{person}{Gustav~Eje Henter}.} \bibinfo{year}{2021}\natexlab{}.
\newblock \showarticletitle{HEMVIP: Human evaluation of multiple videos in
  parallel}. In \bibinfo{booktitle}{\emph{Proceedings of the International
  Conference on Multimodal Interaction}}.
\newblock


\bibitem[\protect\citeauthoryear{Kendall}{Kendall}{1938}]%
        {kendall1938}
\bibfield{author}{\bibinfo{person}{Maurice~G Kendall}.}
  \bibinfo{year}{1938}\natexlab{}.
\newblock \showarticletitle{A New Measure of Rank Correlation}.
\newblock \bibinfo{journal}{\emph{Biometrika}} \bibinfo{volume}{30},
  \bibinfo{number}{1} (\bibinfo{year}{1938}), \bibinfo{pages}{81--93}.
\newblock
\urldef\tempurl%
\url{https://doi.org/10/ch8zq6}
\showDOI{\tempurl}


\bibitem[\protect\citeauthoryear{Kucherenko, Hasegawa, Kaneko, Henter, and
  Kjellstr{\"o}m}{Kucherenko et~al\mbox{.}}{2021a}]%
        {kucherenko2021moving}
\bibfield{author}{\bibinfo{person}{Taras Kucherenko}, \bibinfo{person}{Dai
  Hasegawa}, \bibinfo{person}{Naoshi Kaneko}, \bibinfo{person}{Gustav~Eje
  Henter}, {and} \bibinfo{person}{Hedvig Kjellstr{\"o}m}.}
  \bibinfo{year}{2021}\natexlab{a}.
\newblock \showarticletitle{Moving fast and slow: Analysis of representations
  and post-processing in speech-driven automatic gesture generation}.
\newblock \bibinfo{journal}{\emph{Int. J. Hum. Comput. Interact.}}
  (\bibinfo{year}{2021}).
\newblock
\urldef\tempurl%
\url{https://doi.org/10.1080/10447318.2021.1883883}
\showDOI{\tempurl}


\bibitem[\protect\citeauthoryear{Kucherenko, Jonell, van Waveren, Henter,
  Alexandersson, Leite, and Kjellstr{\"o}m}{Kucherenko et~al\mbox{.}}{2020}]%
        {kucherenko2020gesticulator}
\bibfield{author}{\bibinfo{person}{Taras Kucherenko}, \bibinfo{person}{Patrik
  Jonell}, \bibinfo{person}{Sanne van Waveren}, \bibinfo{person}{Gustav~Eje
  Henter}, \bibinfo{person}{Simon Alexandersson}, \bibinfo{person}{Iolanda
  Leite}, {and} \bibinfo{person}{Hedvig Kjellstr{\"o}m}.}
  \bibinfo{year}{2020}\natexlab{}.
\newblock \showarticletitle{Gesticulator: A framework for semantically-aware
  speech-driven gesture generation}. In \bibinfo{booktitle}{\emph{Proceedings
  of the International Conference on Multimodal Interaction}}.
  \bibinfo{pages}{242--250}.
\newblock


\bibitem[\protect\citeauthoryear{Kucherenko, Jonell, Yoon, Wolfert, and
  Henter}{Kucherenko et~al\mbox{.}}{2021b}]%
        {kucherenko2021large}
\bibfield{author}{\bibinfo{person}{Taras Kucherenko}, \bibinfo{person}{Patrik
  Jonell}, \bibinfo{person}{Youngwoo Yoon}, \bibinfo{person}{Pieter Wolfert},
  {and} \bibinfo{person}{Gustav~Eje Henter}.} \bibinfo{year}{2021}\natexlab{b}.
\newblock \showarticletitle{A large, crowdsourced evaluation of gesture
  generation systems on common data: The GENEA Challenge 2020}. In
  \bibinfo{booktitle}{\emph{26th International Conference on Intelligent User
  Interfaces}}. \bibinfo{pages}{11--21}.
\newblock


\bibitem[\protect\citeauthoryear{Kuznetsova, Brockhoff, and
  Christensen}{Kuznetsova et~al\mbox{.}}{2017}]%
        {kuznetsova2017}
\bibfield{author}{\bibinfo{person}{Alexandra Kuznetsova},
  \bibinfo{person}{Per~B Brockhoff}, {and} \bibinfo{person}{Rune H~B
  Christensen}.} \bibinfo{year}{2017}\natexlab{}.
\newblock \showarticletitle{{{lmerTest Package}}: Tests in Linear Mixed Effects
  Models}.
\newblock \bibinfo{journal}{\emph{Journal of Statistical Software}}
  \bibinfo{volume}{82}, \bibinfo{number}{13} (\bibinfo{year}{2017}),
  \bibinfo{pages}{1--26}.
\newblock
\urldef\tempurl%
\url{https://doi.org/10/dg3k}
\showDOI{\tempurl}


\bibitem[\protect\citeauthoryear{Liang, Zou, and Yu}{Liang
  et~al\mbox{.}}{2020}]%
        {liang2020beyond}
\bibfield{author}{\bibinfo{person}{Weixin Liang}, \bibinfo{person}{James Zou},
  {and} \bibinfo{person}{Zhou Yu}.} \bibinfo{year}{2020}\natexlab{}.
\newblock \showarticletitle{Beyond user self-reported likert scale ratings: A
  comparison model for automatic dialog evaluation}.
\newblock \bibinfo{journal}{\emph{arXiv preprint arXiv:2005.10716}}
  (\bibinfo{year}{2020}).
\newblock


\bibitem[\protect\citeauthoryear{Lucca and Wilbourn}{Lucca and
  Wilbourn}{2018}]%
        {lucca2018communicating}
\bibfield{author}{\bibinfo{person}{Kelsey Lucca} {and}
  \bibinfo{person}{Makeba~Parramore Wilbourn}.}
  \bibinfo{year}{2018}\natexlab{}.
\newblock \showarticletitle{Communicating to learn: Infants’ pointing
  gestures result in optimal learning}.
\newblock \bibinfo{journal}{\emph{Child development}} \bibinfo{volume}{89},
  \bibinfo{number}{3} (\bibinfo{year}{2018}), \bibinfo{pages}{941--960}.
\newblock


\bibitem[\protect\citeauthoryear{Martinez, Yannakakis, and Hallam}{Martinez
  et~al\mbox{.}}{2014}]%
        {martinez2014}
\bibfield{author}{\bibinfo{person}{Hector Martinez}, \bibinfo{person}{Georgios
  Yannakakis}, {and} \bibinfo{person}{John Hallam}.}
  \bibinfo{year}{2014}\natexlab{}.
\newblock \showarticletitle{Don't Classify Ratings of Affect; Rank Them!}
\newblock \bibinfo{journal}{\emph{IEEE Transactions on Affective Computing}}
  \bibinfo{volume}{3045}, \bibinfo{number}{c} (\bibinfo{year}{2014}),
  \bibinfo{pages}{1--1}.
\newblock
\urldef\tempurl%
\url{https://doi.org/10/f6pnzt}
\showDOI{\tempurl}


\bibitem[\protect\citeauthoryear{McGraw and Wong}{McGraw and Wong}{1996}]%
        {mcgraw1996}
\bibfield{author}{\bibinfo{person}{Kenneth~O McGraw} {and} \bibinfo{person}{S~P
  Wong}.} \bibinfo{year}{1996}\natexlab{}.
\newblock \showarticletitle{Forming Inferences about Some Intraclass
  Correlation Coefficients}.
\newblock \bibinfo{journal}{\emph{Psychological Methods}} \bibinfo{volume}{1},
  \bibinfo{number}{1} (\bibinfo{year}{1996}), \bibinfo{pages}{30--46}.
\newblock
\urldef\tempurl%
\url{https://doi.org/10/br5ffs}
\showDOI{\tempurl}


\bibitem[\protect\citeauthoryear{McNeill}{McNeill}{1992}]%
        {mcneill1992hand}
\bibfield{author}{\bibinfo{person}{David McNeill}.}
  \bibinfo{year}{1992}\natexlab{}.
\newblock \bibinfo{booktitle}{\emph{Hand and mind: What gestures reveal about
  thought}}.
\newblock \bibinfo{publisher}{University of Chicago press}.
\newblock


\bibitem[\protect\citeauthoryear{Mueser, Grau, Sussman, and Rosen}{Mueser
  et~al\mbox{.}}{1984}]%
        {mueser1984you}
\bibfield{author}{\bibinfo{person}{Kim~T Mueser}, \bibinfo{person}{Barry~W
  Grau}, \bibinfo{person}{Steve Sussman}, {and} \bibinfo{person}{Alexander~J
  Rosen}.} \bibinfo{year}{1984}\natexlab{}.
\newblock \showarticletitle{You're only as pretty as you feel: facial
  expression as a determinant of physical attractiveness.}
\newblock \bibinfo{journal}{\emph{Journal of Personality and Social
  Psychology}} \bibinfo{volume}{46}, \bibinfo{number}{2}
  (\bibinfo{year}{1984}), \bibinfo{pages}{469}.
\newblock


\bibitem[\protect\citeauthoryear{P{\'e}rez-Mayos, Farr{\'u}s, and
  Adell}{P{\'e}rez-Mayos et~al\mbox{.}}{2019}]%
        {perez2019part}
\bibfield{author}{\bibinfo{person}{Laura P{\'e}rez-Mayos},
  \bibinfo{person}{Mireia Farr{\'u}s}, {and} \bibinfo{person}{Jordi Adell}.}
  \bibinfo{year}{2019}\natexlab{}.
\newblock \showarticletitle{Part-of-speech and prosody-based approaches for
  robot speech and gesture synchronization}.
\newblock \bibinfo{journal}{\emph{Journal of Intelligent \& Robotic Systems}}
  (\bibinfo{year}{2019}), \bibinfo{pages}{1--11}.
\newblock


\bibitem[\protect\citeauthoryear{Phelps, Naeger, Courtier, Lambert, Marcovici,
  Villanueva-Meyer, and MacKenzie}{Phelps et~al\mbox{.}}{2015}]%
        {phelps2015pairwise}
\bibfield{author}{\bibinfo{person}{Andrew~S Phelps}, \bibinfo{person}{David~M
  Naeger}, \bibinfo{person}{Jesse~L Courtier}, \bibinfo{person}{Jack~W
  Lambert}, \bibinfo{person}{Peter~A Marcovici}, \bibinfo{person}{Javier~E
  Villanueva-Meyer}, {and} \bibinfo{person}{John~D MacKenzie}.}
  \bibinfo{year}{2015}\natexlab{}.
\newblock \showarticletitle{Pairwise comparison versus Likert scale for
  biomedical image assessment}.
\newblock \bibinfo{journal}{\emph{American Journal of Roentgenology}}
  \bibinfo{volume}{204}, \bibinfo{number}{1} (\bibinfo{year}{2015}),
  \bibinfo{pages}{8--14}.
\newblock


\bibitem[\protect\citeauthoryear{Prieto~Vives, Igualada~P{\'e}rez, and
  Esteve~Gibert}{Prieto~Vives et~al\mbox{.}}{2017}]%
        {prieto2017beat}
\bibfield{author}{\bibinfo{person}{Pilar Prieto~Vives},
  \bibinfo{person}{Alfonso Igualada~P{\'e}rez}, {and}
  \bibinfo{person}{N{\'u}ria Esteve~Gibert}.} \bibinfo{year}{2017}\natexlab{}.
\newblock \showarticletitle{Beat gestures improve word recall in 3-to
  5-year-old children}.
\newblock \bibinfo{journal}{\emph{Journal of Experimental Child Psychology.
  2017 Apr; 156: 99-112}} (\bibinfo{year}{2017}).
\newblock


\bibitem[\protect\citeauthoryear{Rhemtulla, link will open in a new~window
  {Link to external site}, {Brosseau-Liard}, and Savalei}{Rhemtulla
  et~al\mbox{.}}{2012}]%
        {rhemtulla2012}
\bibfield{author}{\bibinfo{person}{Mijke Rhemtulla}, \bibinfo{person}{this link
  will open in a new~window {Link to external site}},
  \bibinfo{person}{Patricia~{\'E} {Brosseau-Liard}}, {and}
  \bibinfo{person}{Victoria Savalei}.} \bibinfo{year}{2012}\natexlab{}.
\newblock \showarticletitle{When Can Categorical Variables Be Treated as
  Continuous? {{A}} Comparison of Robust Continuous and Categorical {{SEM}}
  Estimation Methods under Suboptimal Conditions}.
\newblock \bibinfo{journal}{\emph{Psychological Methods}} \bibinfo{volume}{17},
  \bibinfo{number}{3} (\bibinfo{year}{2012}), \bibinfo{pages}{354--373}.
\newblock
\urldef\tempurl%
\url{https://doi.org/10/f395ws}
\showDOI{\tempurl}


\bibitem[\protect\citeauthoryear{Salem, Eyssel, Rohlfing, Kopp, and
  Joublin}{Salem et~al\mbox{.}}{2013}]%
        {salem2013err}
\bibfield{author}{\bibinfo{person}{Maha Salem}, \bibinfo{person}{Friederike
  Eyssel}, \bibinfo{person}{Katharina Rohlfing}, \bibinfo{person}{Stefan Kopp},
  {and} \bibinfo{person}{Frank Joublin}.} \bibinfo{year}{2013}\natexlab{}.
\newblock \showarticletitle{To err is human (-like): Effects of robot gesture
  on perceived anthropomorphism and likability}.
\newblock \bibinfo{journal}{\emph{International Journal of Social Robotics}}
  \bibinfo{volume}{5}, \bibinfo{number}{3} (\bibinfo{year}{2013}),
  \bibinfo{pages}{313--323}.
\newblock


\bibitem[\protect\citeauthoryear{Schoeffler, Bartoschek, St{\"o}ter, Roess,
  Westphal, Edler, and Herre}{Schoeffler et~al\mbox{.}}{2018}]%
        {schoeffler2018webmushra}
\bibfield{author}{\bibinfo{person}{Michael Schoeffler}, \bibinfo{person}{Sarah
  Bartoschek}, \bibinfo{person}{Fabian-Robert St{\"o}ter},
  \bibinfo{person}{Marlene Roess}, \bibinfo{person}{Susanne Westphal},
  \bibinfo{person}{Bernd Edler}, {and} \bibinfo{person}{J{\"u}rgen Herre}.}
  \bibinfo{year}{2018}\natexlab{}.
\newblock \showarticletitle{webMUSHRA—A comprehensive framework for web-based
  listening tests}.
\newblock \bibinfo{journal}{\emph{Journal of Open Research Software}}
  \bibinfo{volume}{6}, \bibinfo{number}{1} (\bibinfo{year}{2018}).
\newblock


\bibitem[\protect\citeauthoryear{Simms, Zelazny, Williams, and Bernstein}{Simms
  et~al\mbox{.}}{2019}]%
        {simms2019}
\bibfield{author}{\bibinfo{person}{Leonard~J Simms}, \bibinfo{person}{Kerry
  Zelazny}, \bibinfo{person}{Trevor~F Williams}, {and} \bibinfo{person}{Lee
  Bernstein}.} \bibinfo{year}{2019}\natexlab{}.
\newblock \showarticletitle{Does the Number of Response Options Matter?
  {{Psychometric}} Perspectives Using Personality Questionnaire Data}.
\newblock \bibinfo{journal}{\emph{Psychological Assessment}}
  \bibinfo{volume}{31}, \bibinfo{number}{4} (\bibinfo{year}{2019}),
  \bibinfo{pages}{557--566}.
\newblock
\urldef\tempurl%
\url{https://doi.org/10/gfxv4h}
\showDOI{\tempurl}


\bibitem[\protect\citeauthoryear{Sung and Wu}{Sung and Wu}{2018}]%
        {sung2018visual}
\bibfield{author}{\bibinfo{person}{Yao-Ting Sung} {and}
  \bibinfo{person}{Jeng-Shin Wu}.} \bibinfo{year}{2018}\natexlab{}.
\newblock \showarticletitle{The Visual Analogue Scale for Rating, Ranking and
  Paired-Comparison (VAS-RRP): a new technique for psychological measurement}.
\newblock \bibinfo{journal}{\emph{Behavior research methods}}
  \bibinfo{volume}{50}, \bibinfo{number}{4} (\bibinfo{year}{2018}),
  \bibinfo{pages}{1694--1715}.
\newblock


\bibitem[\protect\citeauthoryear{Wagner, Beskow, Betz, Edlund, Gustafson,
  Henter, Le~Maguer, Malisz, Sz{\'e}kely, T\r{a}nnander, and Vo{\ss}e}{Wagner
  et~al\mbox{.}}{2019}]%
        {wagner2019speech}
\bibfield{author}{\bibinfo{person}{Petra Wagner}, \bibinfo{person}{Jonas
  Beskow}, \bibinfo{person}{Simon Betz}, \bibinfo{person}{Jens Edlund},
  \bibinfo{person}{Joakim Gustafson}, \bibinfo{person}{Gustav~Eje Henter},
  \bibinfo{person}{S{\'e}bastien Le~Maguer}, \bibinfo{person}{Zofia Malisz},
  \bibinfo{person}{{\'E}va Sz{\'e}kely}, \bibinfo{person}{Christina
  T\r{a}nnander}, {and} \bibinfo{person}{Jana Vo{\ss}e}.}
  \bibinfo{year}{2019}\natexlab{}.
\newblock \showarticletitle{Speech synthesis evaluation -- State-of-the-art
  assessment and suggestion for a novel research program}. In
  \bibinfo{booktitle}{\emph{Proc. SSW}}, Vol.~\bibinfo{volume}{10}.
  \bibinfo{publisher}{ISCA}, \bibinfo{address}{Vienna, Austria},
  \bibinfo{pages}{105--110}.
\newblock
\urldef\tempurl%
\url{https://doi.org/10.21437/SSW.2019-19}
\showDOI{\tempurl}


\bibitem[\protect\citeauthoryear{Weijters, Cabooter, and Schillewaert}{Weijters
  et~al\mbox{.}}{2010}]%
        {weijters2010effect}
\bibfield{author}{\bibinfo{person}{Bert Weijters}, \bibinfo{person}{Elke
  Cabooter}, {and} \bibinfo{person}{Niels Schillewaert}.}
  \bibinfo{year}{2010}\natexlab{}.
\newblock \showarticletitle{The effect of rating scale format on response
  styles: The number of response categories and response category labels}.
\newblock \bibinfo{journal}{\emph{International Journal of Research in
  Marketing}} \bibinfo{volume}{27}, \bibinfo{number}{3} (\bibinfo{year}{2010}),
  \bibinfo{pages}{236--247}.
\newblock


\bibitem[\protect\citeauthoryear{Wolfert, Kucherenko, Kjellstr{\"o}m, and
  Belpaeme}{Wolfert et~al\mbox{.}}{2019}]%
        {wolfert2019should}
\bibfield{author}{\bibinfo{person}{Pieter Wolfert}, \bibinfo{person}{Taras
  Kucherenko}, \bibinfo{person}{Hedvig Kjellstr{\"o}m}, {and}
  \bibinfo{person}{Tony Belpaeme}.} \bibinfo{year}{2019}\natexlab{}.
\newblock \showarticletitle{Should Beat Gestures Be Learned Or Designed?: A
  Benchmarking User Study}. In \bibinfo{booktitle}{\emph{ICDL-EPIROB 2019
  Workshop on Naturalistic Non-Verbal and Affective Human-Robot Interactions}}.
\newblock


\bibitem[\protect\citeauthoryear{Wolfert, Robinson, and Belpaeme}{Wolfert
  et~al\mbox{.}}{[n.d.]}]%
        {wolfert2021review}
\bibfield{author}{\bibinfo{person}{Pieter Wolfert}, \bibinfo{person}{Nicole
  Robinson}, {and} \bibinfo{person}{Tony Belpaeme}.}
  \bibinfo{year}{[n.d.]}\natexlab{}.
\newblock \showarticletitle{A Review of Evaluation Practices of Gesture
  Generation in Embodied Conversational Agents}.
\newblock \bibinfo{journal}{\emph{arXiv preprint arXiv:2101.03769}}
  (\bibinfo{year}{[n.\,d.]}).
\newblock


\bibitem[\protect\citeauthoryear{Yannakakis, Cowie, and Busso}{Yannakakis
  et~al\mbox{.}}{2021}]%
        {yannakakis2021}
\bibfield{author}{\bibinfo{person}{Georgios Yannakakis}, \bibinfo{person}{Roddy
  Cowie}, {and} \bibinfo{person}{Carlos Busso}.}
  \bibinfo{year}{2021}\natexlab{}.
\newblock \showarticletitle{The Ordinal Nature of Emotions: An Emerging
  Approach}.
\newblock \bibinfo{journal}{\emph{IEEE Transactions on Affective Computing}}
  \bibinfo{volume}{12}, \bibinfo{number}{1} (\bibinfo{year}{2021}),
  \bibinfo{pages}{16--35}.
\newblock
\urldef\tempurl%
\url{https://doi.org/10.1109/TAFFC.2018.2879512}
\showDOI{\tempurl}


\bibitem[\protect\citeauthoryear{Yannakakis and Mart{\'i}nez}{Yannakakis and
  Mart{\'i}nez}{2015}]%
        {yannakakis2015}
\bibfield{author}{\bibinfo{person}{Georgios Yannakakis} {and}
  \bibinfo{person}{Héctor~P. Mart{\'i}nez}.} \bibinfo{year}{2015}\natexlab{}.
\newblock \showarticletitle{Grounding Truth via Ordinal Annotation}. In
  \bibinfo{booktitle}{\emph{2015 {{International Conference}} on {{Affective
  Computing}} and {{Intelligent Interaction}} ({{ACII}})}}.
  \bibinfo{pages}{574--580}.
\newblock
\showISSN{2156-8111}
\urldef\tempurl%
\url{https://doi.org/10/gjp74q}
\showDOI{\tempurl}


\bibitem[\protect\citeauthoryear{Yoon, Cha, Lee, Jang, Lee, Kim, and Lee}{Yoon
  et~al\mbox{.}}{2020}]%
        {yoon2020speech}
\bibfield{author}{\bibinfo{person}{Youngwoo Yoon}, \bibinfo{person}{Bok Cha},
  \bibinfo{person}{Joo-Haeng Lee}, \bibinfo{person}{Minsu Jang},
  \bibinfo{person}{Jaeyeon Lee}, \bibinfo{person}{Jaehong Kim}, {and}
  \bibinfo{person}{Geehyuk Lee}.} \bibinfo{year}{2020}\natexlab{}.
\newblock \showarticletitle{Speech gesture generation from the trimodal context
  of text, audio, and speaker identity}.
\newblock \bibinfo{journal}{\emph{ACM Transactions on Graphics (TOG)}}
  \bibinfo{volume}{39}, \bibinfo{number}{6} (\bibinfo{year}{2020}),
  \bibinfo{pages}{1--16}.
\newblock


\end{thebibliography}

\end{document}